\documentclass[year=23]{fmcad}
\ifCLASSINFOpdf
\else
\fi
\usepackage{graphicx}
\usepackage{booktabs}
\usepackage{multirow}
\usepackage{caption}
\usepackage{subcaption}
\usepackage{amssymb}
\usepackage{color}
\usepackage[linesnumbered,ruled,vlined]{algorithm2e}
\newtheorem{definition}{Definition}
\newtheorem{example}{Example}
\begin{document}
\title{Local Search For SMT On Linear and Multi-linear Real Arithmetic}

\author{\IEEEauthorblockN{Bohan Li \orcid{0000-0003-1356-6057}, Shaowei Cai\orcid{0000-0003-1730-6922}\IEEEauthorrefmark{1}}
\IEEEauthorblockA{\textit{State Key Laboratory of Computer Science, Institute of Software, Chinese Academy of Sciences }
}
\IEEEauthorblockA{\textit{School of Computer Science and Technology, University of Chinese Academy of Sciences}\\ Beijing, China\\
\{libh, caisw\}@ios.ac.cn
}}

% \author{\IEEEauthorblockN{Bohan Li\orcid{0000-0003-1356-6057}}
% \IEEEauthorblockA{\textit{State Key Laboratory of Computer Science}\\
% \textit{Institute of Software, Chinese Academy of Sciences} 
% }
% \IEEEauthorblockA{\textit{School of Computer Science and Technology}\\
% \textit{University of Chinese Academy of Sciences}\\ Beijing, China\\
% libh@ios.ac.cn
% }
% \and
% \IEEEauthorblockN{Shaowei Cai\orcid{0000-0003-1730-6922}\IEEEauthorrefmark{1}}
% \IEEEauthorblockA{\textit{State Key Laboratory of Computer Science}\\
% \textit{Institute of Software, Chinese Academy of Sciences} 
% }
% \IEEEauthorblockA{\textit{School of Computer Science and Technology}\\
% \textit{University of Chinese Academy of Sciences}\\ Beijing, China\\
% caisw@ios.ac.cn
% }
% }


\maketitle

\begin{abstract}
Satisfiability Modulo Theories (SMT) has significant application in various domains. In this paper, we focus on  quantifier-free Satisfiablity Modulo Real Arithmetic, referred to as SMT(RA), including both linear and non-linear real arithmetic theories. As for non-linear real arithmetic theory, we focus on one of its important fragments where the atomic constraints are multi-linear. We propose the first local search algorithm for SMT(RA), called LocalSMT(RA), based on two novel ideas. First, an interval-based operator is proposed to cooperate with the traditional local search operator by considering the interval information. Moreover, we propose a tie-breaking mechanism to further evaluate the operations when the operations are indistinguishable according to the score function. Experiments are conducted to evaluate LocalSMT(RA) on benchmarks from SMT-LIB. The results show that LocalSMT(RA) is competitive with the state-of-the-art SMT solvers, and performs particularly well on multi-linear instances.
\end{abstract}


% \fi
\IEEEpeerreviewmaketitle

\section{Introduction}
Satisfiability Modulo Theories (SMT) is the  problem of checking the satisfiability of a first order logic formula with respect to certain background theories.
It has been applied in various areas, including program verification and termination analysis~\cite{lahiri2008back,bjorner2012program}, symbolic execution~\cite{baldoni2018survey} and test-case generation~\cite{peleska2011automated}, etc.

In this paper, we focus on the theory of quantifier-free real arithmetic, consisting of  atomic constraints in the form of polynomial equalities or inequalities over real variables.
The theory can be divided into two categories, namely {\it linear real arithmetic} (LRA) and {\it non-linear real arithmetic} (NRA), based on whether the arithmetic atomic constraints are linear or not.
As for NRA, this paper concerns one of its important fragments where the atomic constraints are multi-linear.
The SMT problem with the background theory of LRA and NRA is to determine the satisfiability of the Boolean combination of respective atomic constraints, referred to as SMT(LRA) and SMT(NRA).
In general, we refer to the SMT problem on the theory of real arithmetic as SMT(RA).

\subsection{Related Work}
%Much effort has been devoted to solve SMT on real arithmetic.
The mainstream approach for solving SMT(RA) is the {\it lazy} approach~\cite{sebastiani2007lazy,barrett2018satisfiability}, also known as DPLL(T)~\cite{nieuwenhuis2006solving}, which relies on the interaction of a SAT solver with a theory solver.
Most state-of-the-art SMT solvers supporting the theory of real arithmetic are mainly based on the {\it lazy} approach, including Z3~\cite{de2008z3}, Yices2~\cite{dutertre2014yices}, SMT-RAT~\cite{corzilius2012smt}, cvc5~\cite{barbosa2022cvc5}, OpenSMT~\cite{bruttomesso2010opensmt} and MathSAT5~\cite{cimatti2013mathsat5}.
In the DPLL(T) framework, the SMT formula is abstracted into a Boolean formula by replacing arithmetic atomic constraints with fresh Boolean variables. A SAT solver is employed to reason about the Boolean structure, while a theory solver is invoked to receive the set of theory constraints determined by the SAT solver, and solve the conjunction of these theory constraints, including consistency checking of the assignments and theory-based deduction.

The efforts in the {\it lazy} approach are mainly devoted to designing effective decision procedures, serving as theory solvers to deal with the conjunction of theory constraints.
The core reasoning module for LRA  integrated in DPLL(T) is a variant of the {\it simplex} algorithm dedicated for SMT solving, proposed in~\cite{dutertre2006fast}.
Another approach for solving LRA constraint systems is the {\it Fourier-Motzkin} variable elimination~\cite{bjorner2010linear}, which often shows worse performance than the {\it simplex} algorithm.

As for non-linear real arithmetic, the {\it cylindrical algebraic decomposition} (CAD)~\cite{collins1975quantifier} is the most widely used decision procedure, and CAD is adapted and embedded as theory solver in the SMT-RAT solver~\cite{corzilius2012smt} with improvement since~\cite{loup2013symbiosis}.
Other well-known methods use Gr{\"o}bner bases~\cite{junges2013grobner} or the realization of sign conditions~\cite{basu2014algorithms}.
Incomplete methods include a theory solver~\cite{corzilius2011virtual}  based on virtual substitution~\cite{weispfenning1997quantifier}, and techniques based on interval constraint propagation~\cite{van1997solving} proposed in~\cite{gao2013dreal,schupp2013interval}.

Moreover, Constructing Satisfiability (MCSat) calculus is also an efficient framework for solving SMT(RA). It was proposed in~\cite{jovanovic2013design} for solving SMT(LRA), and an elegant variation of CAD method is instantiated in the model-constructing satisfiability calculus framework of Z3~\cite{jovanovic2013solving} for solving SMT(NRA).

Local search  is an incomplete method playing a significant part in many  combinatorial problems   ~\cite{hoos2004stochastic}.
It has been successfully applied to the Boolean Satisfiability (SAT) problem   ~\cite{LiL12,BalintS12,cai2013local,cai2015ccanr,Biere17} and can rival CDCL solvers on certain types of instances.

Local search for SMT, however, has only received very little amount of attention.
The idea of integrating local search solvers with theory solvers has been previously explored to solve SMT(LRA), where a local search SAT solver WalkSAT is used to solve the Boolean skeleton of the SMT formula~\cite{griggio2011stochastic}.
 A local search solver called LocalSMT was recently employed to SMT on integer arithmetic~\cite{cai2022local,cai2023local}.
 Moreover, local search algorithm has been applied to Bit-vectors ~\cite{DBLP:journals/fmsd/NiemetzPB17,DBLP:conf/fmcad/NiemetzP20,DBLP:conf/aaai/FrohlichBWH15}.
 However, we are not aware of any local search algorithm for SMT on real arithmetic.

\subsection{Contributions}

In this paper, for the first time, we design a local search algorithm for SMT(RA), namely LocalSMT(RA),  based on the following novel strategies.
Note that LocalSMT(RA) is implemented as a fragment of LocalSMT~\cite{cai2023local}, which is our local search solver dedicated for SMT.

First, we propose the  {\it interval-based} operator to enhance the conventional local search operator by taking interval information into account.
Specifically, we observe that assigning the real-value variable to any value in a given interval  would make the same amount of currently falsified clauses become satisfied. 
Hence, the {\it interval-based} operator evaluates multiple values inside the interval as the potential value of the operation, rather than only assign it to a fixed value (e.g. the threshold value to satisfy a constraint).

Moreover, we observe that there frequently exist multiple  operations with the same best score when performing local search, and thus a tie-breaking mechanism is proposed to further distinguish these operations.
% by preferring the operation that assigns a variable to a value with small denominator.
%By putting these together, we proposed the first local search algorithm for SMT(RA) called LocalSMT(RA).

Experiments are conducted to evaluate LocalSMT(RA) on 2 benchmark sets, namely SMT(LRA) and SMT(NRA) benchmarks from SMT-LIB.
Note that unsatisfiable instances are excluded, and we only consider multi-linear instances from SMT(NRA) benchmark.
We compare LocalSMT(RA) with the top 4 SMT solvers in the relevant logics (QF\_LRA, QF\_NRA) according to the SMT-COMP 2022\footnote{https://smt-comp.github.io/2022/}, excluding the portfolio and derived solvers.
Specifically, as for SMT(LRA), we compare LocalSMT(RA) with OpenSMT, Yices2, cvc5 and Z3, while for SMT(NRA), the competitors are  Z3, cvc5, Yices2 and SMT-RAT.
Experimental results show that LocalSMT(RA) is competitive and complementary with state-of-the-art SMT solvers, especially on multi-linear instances.
Moreover, the ablation experiment confirms the effectiveness of our proposed novel strategies.

Note that multi-linear instances are comparatively difficult to solve by existing solvers. For example, Z3, perhaps the best solver for satisfiable SMT(NRA) instances according to SMT-COMP 2022, can solve 90.5\% QF(NRA) instances, while it can only solve 77.5\% multi-linear instances. 
However, multi-linear instances are suitable for local search, since without high order terms, the operation can be  efficiently calculated.

\subsection{Paper Organization}
In section \ref{pre}, preliminary knowledge is introduced.
% In section \ref{review}, we review the first local search algorithm for SMT(LIA), whose main features are adopted in our algorithm.
In section \ref{interval op}, we propose a novel {\it interval-based operator} to enrich the traditional operator by considering the interval information.
In section \ref{pick}, a {\it tie-breaking mechanism} is proposed to distinguish multiple operations with the same best score.
Based on the two novel strategies, our local search for SMT(RA) is proposed in section \ref{algorithm}.
Experiments are presented in section \ref{experiments}.
Conclusion and future work are given in section \ref{conclusion}.

\section{Preliminaries}
\label{pre}

\subsection{Basic Definitions}
A $monomial$ is an expression of the form $x_1^{e_1}...x_m^{e_m}$ where $m>0$, $x_i$ are variables and $e_i$ are exponents,
$e_i>0$ for all $i\in\{1...m\}$, and $x_i\not=x_j$ for all $i,j\in\{1...m\},i\not=j$.
A monomial is linear if $m=1$ and $e_1=1$.

A $polynomial$ is a linear combination of monomials, that is, an arithmetic expression of the form $\sum_{i} a_i m_i$ where $a_i$ are coefficients and $m_i$ are monomials.
If all its monomials are linear in a polynomial, indicating that the $polynomial$ can be written as $\sum_i a_ix_i$, then it is $linear$, and otherwise it is {\it non-linear}.
A special case of non-linear polynomial is $multi-linear$ polynomial, where the highest exponent for all variables is 1, indicating that each monomial is in the form of $x_1...x_m$. 

\begin{definition}
    The atomic constraints of the theory of real arithmetic are polynomial inequalities and equalities, in the form of $\sum_{i} a_i m_i\bowtie k$, where $ \bowtie\in \{=,\le,<,\ge,>\}$, $m_i$ are monomials consisting of real-valued variables, $k$ and $a_i$ are rational constants.
\end{definition}

The formulas of the SMT problem on the theory of real arithmetic, denoted as SMT(RA), are Boolean combinations of atomic constraints and propositional variables, where the sets of real-valued variables and propositional variables are denoted as $X$ and $P$.
The SMT problem on the theory of linear real arithmetic (LRA) and non-linear arithmetic (NRA) are denoted as SMT(LRA) and SMT(NRA), respectively.
As for NRA, this paper only considers one important fragment where the polynomials in atomic constraints are multi-linear, denoted as {\it MRA} in this paper.

\begin{example}
Let $X=\{x_1,x_2,x_3,x_4,x_5\}$ and $P=\{p_1,p_2\}$ denote the sets of integer-valued and propositional variables respectively.
A typical SMT(LRA) formula $F_{LRA}$ and SMT(MRA) formula $F_{MRA}$ is shown as follows:

$F_{LRA}$: $(p_1\vee (x_1+2x_2\le2) \;)\wedge(p_2\vee (3x_3+4x_4=2)\vee(-x_2-x_3< 3)\;) $

$F_{MRA}$: $(p_1\vee (x_1 x_2\le2) \;)\wedge(p_2\vee (3x_3x_4+4x_4=2)\vee(-x_2-x_3< 3)\;) $
\end{example}

In the theory of real arithmetic, a positive, infinitesimal real number is denoted as $\delta$.

A literal is an atomic constraint or a propositional variable, or their negation.
A $clause$ is the disjunction of a set of literals, and a formula in {\it conjunctive normal form} (CNF) is the conjunction of a set of clauses.
For an SMT(RA) formula $F$, an assignment $\alpha$ is a  mapping $X\rightarrow R$ and $P\rightarrow \{false,true\}$, and $\alpha(x)$ denotes the value of a variable $x$ under $\alpha$. 
A {\it complete assignment} is a mapping which assigns to each variable a value.
A literal is true if it evaluates to true under the given assignment, and false otherwise.
A clause is {\it satisfied} if it has at least one true literal, and {\it falsified} if all literals in the clause are false.
A complete assignment is a {\it solution} to an SMT(RA) formula iff it satisfies all the clauses.

\subsection{Local Search}

When local search is performed on the SMT problem, the search space is comprised of all complete assignments, each of which represents a candidate solution.
Typically, a local search algorithm begins with a complete assignment and repeatedly updates it by modifying the value of variables in order to find a {\it solution}.

Given a formula $F$, the {\it cost} of an assignment $\alpha$, denoted as $cost(\alpha)$, is the number of falsified clauses under $\alpha$. In dynamic local search algorithms which use clause weighting techniques~\cite{ThorntonPBF04,cai2013local}, $cost(\alpha)$ denotes the total weight of all falsified clauses under an assignment $\alpha$ (The weight is computed according to the PAWS scheme which will be described in detail in Section \ref{sls in mode}).

A  key component of a local search algorithm is the set of {\it operators}, which define how to modify the current solution.
When an operator is instantiated by specifying the variable to operate and the value to assign, an {\it operation} is obtained.
The operation to assign variable $x$ to value $v$ is denoted as $op(x,v)$.
The critical move operator for SMT on linear integer arithmetic proposed in~\cite{cai2022local} is defined as follows.

%In contrast to the Boolean mode where the {\it flip} operator is adopted to modify a Boolean variable to the opposite of its current value, in integer mode, an operator called {\it critical move} working on the literal level is proposed.

\begin{definition}
The critical move operator, denoted as  $cm(x,\ell)$, assigns an integer variable $x$ to the threshold value  making literal $\ell$ true, where $\ell$ is a falsified literal containing $x$.
\end{definition}

For example, given a falsified literal $\ell=(x+1\le0)$ where $x$ is currently assigned to 0, the corresponding  operation $cm(x,\ell)$ will assign $x$ to $-1$.

Local search algorithms usually choose an operation among candidate operations according to some scoring function.
% which can be a property of operations or any mathematical expression of the properties.
Given a formula and an assignment $\alpha$, the most commonly used scoring function of an operation $op$ is defined as $$score(op)=cost(\alpha')-cost(\alpha)$$ where $\alpha'$ is the resulting assignment by applying $op$ to $\alpha$.
An operation $op$ is said to be $decreasing$ if $score(op)>0$.

Another property used for evaluating an operation is its {\it make value}.

\begin{definition}
Given an operation $op$, the make value of $op$, denoted as $make(op)$, is the number of falsified clauses that would become satisfied after performing $op$.
\end{definition}

\section{Interval-based Operation}
\label{interval op}

{\it Critical move}  satisfies falsified clauses by modifying one variable in a false literal to make it true.
This operator can still be used in the context of SMT(RA), and it is also used in our algorithm. 
However, an issue of accuracy arises when applying the critical move operator in the context of Real Arithmetic. Recalling that we need to calculate the threshold value for a literal to become true, when solving a strict inequality, there is no threshold value. Instead, the value depends on what accuracy we intend to maintain. In this section, we propose an operator for SMT(RA), which considers the interval information and is more flexible than critical move.
%First, for a real-valued variable appearing in strict inequalities, keeping assigning it to the threshold requires high accuracy, bringing burden for the calculation of operation.

\subsection{Satisfying Domain}
An important fact on linear or multi-linear inequality of real-value variables is that, when all variables but one in the inequality  are fixed, there is a domain for the remaining variable whose coefficient is not 0, such that assigning the variable to any value in the domain makes the inequality hold.
Thus, given a falsified literal $\ell$ in the form of an atomic constraint and a variable $x$ in it, it can be satisfied by assigning $x$ to any value in the corresponding domain, called {\it Satisfying Domain}.
%for literal level, denoted as $SI_{l}(\ell,x)$.
For example, consider a literal $\ell: (x-y> 4)$ where the current assignment is $\alpha=\{x=0, y=0\}$, then obviously assigning $x$ to any value in   $(4,+\infty)$ satisfies the inequality, and thus the {\it Satisfying Domain} is $(4,+\infty)$.

% Taking further steps by calculation on such intervals, 
We further extend the definition of {\it Satisfying Domain} to the clause level, defined as follows.

\begin{definition}
Given an assignment $\alpha$, for a false literal $\ell$ and a variable $x$ appearing in $\ell$, the {\bf satisfying domain of $x$ for literal} $\ell$ is $SD_{l}(x,\ell)=\{v|\ell$ becomes true if assigning $x$ to $v\}$;
for a falsified clause $c$ and a variable $x$ in $c$, the {\bf satisfying domain of $x$ for clause} $c$ is $SD_c(x,c)=\bigcup_{\ell\in c}SD_l(x,\ell) $.

\end{definition}

Since the false literal $\ell$ is in the form of linear or multi-linear inequality, $SD_l(x,\ell)$ is either in the form of $(-\infty,u]$ or $[l,\infty)$.
Thus, as the union of $SD_l(x,\ell)$, $SD_c(x,c)$ may contain $(-\infty,u]$ whose upper bound is defined as $UB(x,c)=u$, or $[l,\infty)$ whose lower bound is defined as $LB(x,c)=l$, or both kinds of intervals.
 For simplicity,  interval $(-\infty,u)$ or $(l,\infty)$ are denoted as $(-\infty,u-\delta]$ or $[l+\delta,\infty)$ respectively.

\begin{example}
    Given a clause $c=\ell_1\vee\ell_2\vee\ell_3=(a-b>4) \vee (2a-b \ge 7) \vee (2a-c \le-5)$ and the current assignment $\alpha=\{a=0,b=0,c=0\}$, for variable $a$, the satisfying domains to the three literals are $SD_l(a,\ell_1)=[4+\delta,\infty)$, $SD_l(a,\ell_2)=[3.5,\infty)$ and $SD_l(a,\ell_3)=(-\infty,-2.5]$ respectively. The {\it Satisfying Domain} to clause $c$ is $SD_c(a,c)=(-\infty,-2.5]\cup[3.5,\infty)$, and the corresponding upper bound and lower bound are $UB(a,c)=-2.5$ and $LB(a,c)=3.5$.
\end{example}

\subsection{Equi-make Intervals}
% Given a variable $x$ in falsified clauses, $I_{m}(x)$ denotes the set of intervals to which assigning $x$ can satisfy the same amount of clauses.
% The procedure to determine $I_{m}$ 

Based on the variables' {\it satisfying domain} to clauses, we observe that operations assigning the variable to any value in a given interval would satisfy the same amount of falsified clauses, that is, they have the same {\it make value}. % and there exist infinite options in such intervals. 
We call such interval as {\it equi-make interval}.

\begin{definition}
Given an SMT(RA) formula $F$ and an assignment $\alpha$ to its variables, for a variable $x$, an  {\bf equi-make interval} is a maximal interval $I$ such that all operations $op(x,v)$ with $v\in I$  have the same make value.
\end{definition}

We can divide $(-\infty, +\infty)$ into several equi-make intervals w.r.t. a variable. 

\begin{example}
Consider a formula $F: c_1\wedge c_2$ where both clauses are falsified under the current assignment and  variable $a$ appears in both clauses.
Suppose $SD_c(a,c_1)=[3,+\infty)$ and $SD_c(a,c_2)=[5,+\infty)$, then we can divide $(-\infty,+\infty)$ into three intervals as $(-\infty, 3)$, $[3,5)$ and $[5,+\infty)$. Operations assigning $a$ to any value in $(-\infty, 3)$ results in a make value of 0, those assigning $a$ to a value in $[3,5)$ results in a make value of 1, while those corresponding to $[5,\infty)$ results in a make value of 2.
\end{example}

% Thus, instead of focusing on calculating the threshold value as in  {\it critical moves}, we focus on finding the good intervals.  This enriches our options for the value of a given variable, as picking any value in such an interval satisfies the same amount of falsified clauses.
Thus, we can enrich the traditional {\it critical move} operator by considering the interval information.
The intuition is to find the equi-make intervals, and then consider multiple values in such interval as the options for future value of operations, rather than only consider the threshold value.

% For each real variable $x$ in falsified clauses, we hope to find such interval $I$ that assigning $x$ to any value $v\in I$ can satisfy the same amount of clauses.
% The procedure to determine the set of such intervals, denoted as $I(x)$, are as follows.

We focus on the variables appearing in at least one falsified clause.
Here we describe a procedure to  partition $(-\infty,+\infty)$ into equi-make intervals for such variables.

\begin{itemize}
    \item First, we go through the falsified clauses. For each  falsified clause $c$, we calculate for each real-valued variable $x$ in $c$  the corresponding {\it satisfying domain} to $c$, $SD_c(x,c)$, as well as the upper bound $UB(x,c)$ and lower bound $LB(x,c)$ if they exist.
    \item Then, for each real-valued variable $x$ appearing in falsified clauses, all its $UBs$  are sorted in the ascending order, while $LBs$  sorted in the descending order.
    After sorting, these bounds are labeled as $UB^1(x),\dots UB^n(x)$ and $LB^1(x)$,$\dots$ $ LB^m(x)$, where $UB^n(x)$ and $LB^m(x)$ denotes the maximum of $UB$ and minimum of $LB$ for $x$ respectively \footnote{Note that $UB^n(x)<LB^m(x)$, since otherwise the current assignment is either in the interval $[LB^m(x),\infty)$ or  $(-\infty,UB^n(x)]$. Suppose the falsified clauses corresponding to the intervals $[LB^m(x),\infty)$ and $(-\infty,UB^n(x)]$ are $c_1$ and $c_2$, if the current assignment is either in the first or second interval, then according to the definition, either $c_1$ or $c_2$ has already been satisfied, which contradicts the definition of $UB^n$ and $LB^m$. }.
For convenience in description, we denote $UB^0(x)=-\infty$ and $LB^0(x)=\infty$.
These bounds are listed in order: $UB^0(x)<UB^1(x)<\dots<UB^n(x)<LB^{m}(x)<\dots<LB^{1}(x)<LB^{0}(x)$.
\item Finally, for each variable $x$, we obtain an interval partition 
$IP(x)=\bigcup_{0<i\le n}\{(UB^{i-1}(x),UB^{i}(x)]\}\cup (UB^n, LB^m) \cup  \bigcup_{0<j\le m}\{[LB^{j}(x),LB^{j-1}(x))\}$
% Note that the make value of each operation assigning the variable to any value in intervals are positive except $(UB^n, LB^m)$, for which make value is 0.

%It is formally stated as follows: Given a real variable $x$ and an interval $I\in IP(x)$, $\forall v_1,v_2\in I$ where $v_1\not =v_2$, $make(op(x,v_1))=make(op(x,v_2))$.

%%%to do : consider the interval (UB^n, LB^m]
\end{itemize}

Formally, given a real variable $x$ and an interval $I$ from $ IP(x)$, $\forall v_1,v_2\in I$, $make(op(x,v_1))=make(op(x,v_2))$.
As a slight abuse of notation, for an interval $I$ from $IP(x)$, we define its {\it make value} as the make value of any operation $op(x,v)$ with $v\in I$. 
Note that all intervals in $IP(x)$ have positive make values except $(UB^n, LB^m)$, whose  make value is 0.

\begin{figure}[ht]
\centering
\centerline{\includegraphics[scale=0.41]{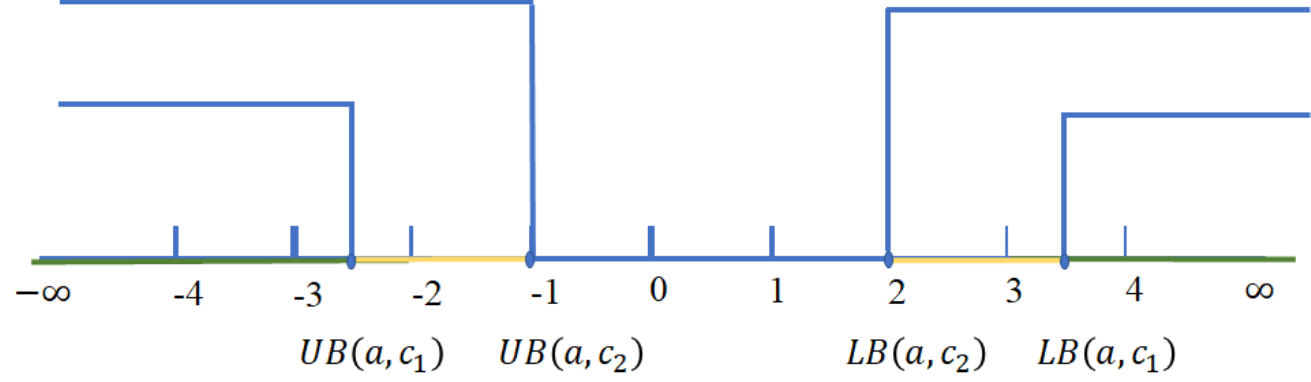}}
\caption{Interval example}
\label{interval}
\end{figure}

\begin{example}
Given a formula $F:c_1\wedge c_2=(a-b > 4\vee 2a-b\ge 7\vee 2a-c\le -5)\wedge (a-c\ge 2\vee a-d\le-1)$ and the current assignment is $\alpha=\{a=0, b=0, c=0\}$.
For variable $a$, $SD_c(a,c_1)=(-\infty,-2.5]\cup[3.5,\infty)$ and $SD_c(a,c_2)=(-\infty,-1]\cup[2,\infty)$.
% And thus, $LB(c_1,a)=3.5$, $UB(c_1,a)=-2.5$, $LB(c_2,a)=2$, and $UB(c_2,a)=-1-\delta$.
Then, $UB^0(a)=-\infty$, $UB^1(a)=UB(a,c_1)=-2.5$, $UB^2(a)=UB(a,c_2)=-1$, $LB^2(a)=LB(a,c_2)=2$, $LB^1(a)=LB(a,c_1)=3.5$ and $LB^0(a)=\infty$.
% Therefore, assigning $a$ to any value in interval $[3.5,\infty)$ and $[2,3.5)$ can satisfy 2 and 1 clauses respectively.

Therefore, interval partition for $x$ is $IP(x)=I_1 \cup I_2 \cup I_3 \cup I_4 \cup I_5$=$(-\infty, -2.5]\cup (-2.5, -1] \cup (-1, 2) \cup [2, 3.5) \cup [3.5, +\infty)$, as shown in Fig \ref{interval}. For these intervals w.r.t. $x$, the make value is 2, 1, 0, 1, 2 respectively.

%assigning $x$ to any value in green intervals can satisfy 2 clauses, while assigning $x$ to any value in yellow intervals can satisfy 1 clause.
\end{example}

\subsection{Candidate Values for Operations}
% Based on the observation and definition above, since for a real-valued variable $x$, 
% assigning it to any value in the interval $(I_u^{i-1}(x),I_u^{i}(x)]$ or interval $[I_l^{i}(x),I_l^{i-1}(x))$ can satisfy the same amount of falsified clauses,
% $(I_u^{i-1}(x),I_u^{i}(x)]$ or interval $[I_l^{i}(x),I_l^{i-1}(x))$
Since assigning a variable $x$ to any value in an equi-make interval  would satisfy the same amount of falsified clauses, after choosing an equi-make interval, we can consider more values in the interval as the option for the future value of operation, rather than only the threshold.

The motivation for the candidate future values is based on the following 3 intuitions: 
First, we have to restrict the value in the given interval.
Moreover, we hope the denominator of corresponding value to be relatively small, in order to avoid exploring complex search space as will be explained in the next section.
Finally, the value should be easy to calculate, in order to improve the efficiency of algorithm.

In this work, we only consider the intervals with a positive make value, and thus the interval $(UB^n, LB^m)$ is omitted. Thus, the interval for consideration is of the form $(UB^{i-1}(x),UB^{i}(x)]$ or $[LB^{i}(x),LB^{i-1}(x))$. For such an interval, we consider the following values for the operation:
\begin{itemize}
    \item Assign $x$ to the threshold $UB^i(x)$ or $LB^i(x)$.
    \item Assign $x$ to the median of the interval, that is $(UB^{i-1}(x)+UB^{i}(x))/2$ or $(LB^{i}(x)+LB^{i-1}(x))/2$ (when one of the bounds is $\infty$ or $-\infty$, the median will not be considered).
    \item if there are integers in the interval $(UB^{i-1}(x),UB^{i}(x))$ or the interval $(LB^{i}(x)$, $LB^{i-1}(x))$, assign $x$ to the largest or smallest integer in the respective open interval; Otherwise, suppose that the open interval can be written as $(\frac{a}{b},\frac{c}{d})$, then assign $x$ to $\frac{a+c}{b+d}$.
\end{itemize}

The first option is the same as critical move, and thus {\it critical move} can be regarded as a special case of our interval-based operator.
The second option is a simple choice among intervals.
The third option aims to find a rational value limited in the open interval with small denominator, and it is easy to calculate. 

Note that it is difficult and even even impractical to compute the operations leading to the largest local decrease in the score, based on the following reasons: 
An operation consists of the variable to modify and the future value to assign. However, the variable can be assigned with arbitrary real number, which is inexhaustible. Although this can be reduced to enumerable intervals, it is still too time-consuming to enumerate all possible operations.  Moreover, a literal $\ell$ contains different variables, and changing one such variable would affect all other literals (not only $\ell$) containing the variable. Thus, to calculate the score of an operation, we need to go through all literals containing the variable to modify, which is time-consuming. These two reasons make it a very time-consuming procedure to compute such an operation leading to the largest local decrease in the score.

Thus, our algorithm does not tempt to compute such an operation, but chooses an operation with good score from sampled candidate operations with those candidate values.
In the future work, we will further enrich the sampled candidate values by considering more random values with small denominator in the interval.

\section{A Tie-breaking Mechanism}
\label{pick}
%In this section, we propose picking heuristics for selecting operation.
We notice that there often exist different operations with the same best $score$ during local search, and thus tie-breaking is also important to guide the search. 
%IN this section, we propose a tie-breaking heuristic.
% by preferring the operations with small denominator.

To confirm our observation, we conduct a pre-experiment on 100 randomly selected instances.
On each instance, we execute a simple local search algorithm which selects an operation with the best $score$ for  10000 iterations, and we count the number of steps where $k$ operations with the same best $score$ are found, denoted as $step(k)$.

\begin{figure}[ht]
\centering
\centerline{\includegraphics[scale=0.5]{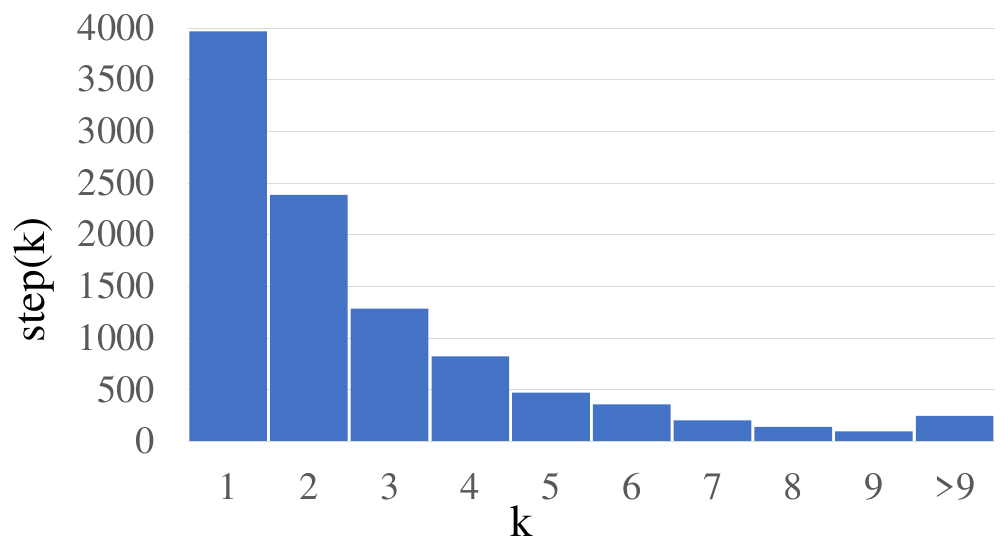}}
\caption{Average $step(k)$ distribution}
\label{step_cnt}
\end{figure}

As shown in Fig. \ref{step_cnt}, the steps where more than one operations have the same best $score$ take up 61.2\% of the total steps.
Thus, a tie-breaking heuristic is required to further distinguish these operations with same best $score$.

First, we consider that assigning real-valued variable to values with large denominator can lead the algorithm to a more complex search space where variables are assigned to real number with extremely large denominators, leading to more complicated computation and possible errors, because the local search solver needs to perform factorization of large numbers and numerical approximation to handle these values during the iteration.
Thus, we prefer operations that assign variable to a value with a small denominator.

Moreover, we consider that assigning variables to values with large absolute value  can lead to the assignment with an extraordinarily large value, deviating the algorithm from finding a possible solution.
Thus, we prefer operations assigning variables to  values with  small absolute value.

Based on the above observation and intuition, we propose a selection rule for picking operations, described as follows.

{\bf Selection Rules}: Select the operation with the greatest $score$, breaking ties by preferring the one assigning the corresponding variable to a value with the smallest denominator.
Further ties are broken by picking the operation assigning the variable with the smallest absolute value.

\section{LocalSMT(RA) Algorithm}
\label{algorithm}

Our local search algorithm adopts a two-mode framework, which switches between  Real mode and Boolean mode. This framework has been used in the local search algorithm LS-LIA for integer arithmetic theories~\cite{cai2022local}.

\subsection{Local search Framework}
\begin{figure}[ht]
\centering
\centerline{\includegraphics[scale=0.45]{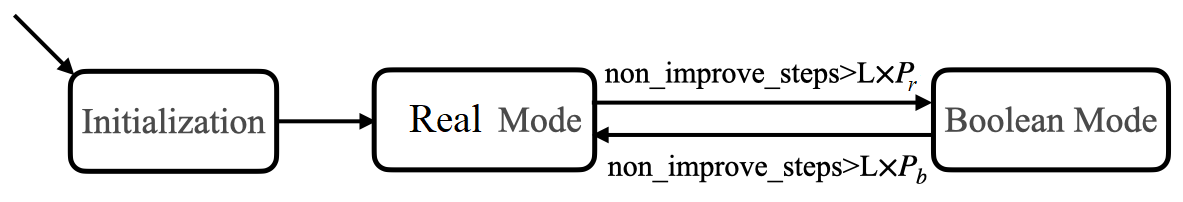}}
\caption{An SMT Local Search Framework}
\label{framework}
\end{figure}

LocalSMT(RA) is running a local search algorithm starting from an arbitrary model with all real-value variables assigned to 0 and all Boolean variables assigned to false.
As depicted in Fig. \ref{framework},  after the initialization, the algorithm switches between Real mode and Boolean mode.
In each mode, an operation on a variable of the corresponding data type is selected to modify the current solution.
 The two modes switches between each other when the number of non-improving steps (denoted as $non\_improve\_steps$) of the current mode reaches a threshold. 
 $non\_improve\_steps$ is increased by one when the algorithm fails to find a better solution, otherwise, it is reset as 0.
 The threshold is set to $L\times P_b$ for the Boolean mode and $L\times P_r$ for the Real mode, where $P_b$ and $P_r$ denote the proportion of Boolean and real-variable literals to all literals in falsified clauses, and $L$ is a parameter.

\subsection{Local Search in Real and Boolean Mode}
\label{sls in mode}

The algorithm for the Real mode of LocalSMT(RA) is described in Algorithm \ref{LS-real}: if the current assignment $\alpha$ satisfies the given formula $F$, then the solution is found (Line 2).
The algorithm tries to find a decreasing {\it interval-based} operation according to the {\bf Selection Rule} (Line 3--4).

If there exists no such decreasing operation, this is an indication that the algorithm falls into the local optimum.
We first update the clause weights according to the probabilistic version of the Pure Additive Weighting Scheme (PAWS) \cite{ThorntonPBF04,cai2013local} (Line 6),
and then randomly sample $K$ interval-based operations into the set $Set_{op}$ (Line 7), where $K$ is a relatively small parameter.
The best operation is picked according to {\bf Selection Rules} among $Set_{op}$ (Line 8).
Note that since the {\it interval-based} operation can satisfy at least one clause, picking the best one among few randomly sampled {\it interval-based} operation  can be regarded as a diversification operation.

The probabilistic version of the PAWS scheme works as follows.
In the beginning, the weight of each clause is initialized as 1.
During the local search process, with probability $1-sp$, the weight of each falsified clause is increased by one, and  with probability $sp$, for each satisfied clause whose weight is greater than 1, the weight is decreased by one.

% As for the Boolean mode focusing on the subformula consisting of Boolean variables, the SMT(RA) problem is reduced to the SAT problem, and our algorithm works in the same way as CCAnr \cite{cai2015ccanr}, the state-of-the-art local search algorithm for SAT, which can be downloaded from its homepage\footnote{http://lcs.ios.ac.cn/~caisw/Code/CCAnr-1.1.zip}.

As for the Boolean mode focusing on the subformula consisting of Boolean variables, LocalSMT(RA) works in the same way as the Boolean mode of LS-LIA.
By putting the Boolean mode and the Real mode together, we propose our local search algorithm for SMT(RA) denoted as LocalSMT(RA).

\begin{algorithm}[!t]
\KwIn{formula $F$ }
\KwOut{ A satisfying assignment $\alpha$ of $F$, or ``UNKNOWN''}
\caption{Real Mode of LocalSMT(RA)}
\label{LS-real}
\While{{\it non\_impr\_steps} $< L\times P_r$}{
\lIf{$\alpha$ satisfies $F$}{return $\alpha$}
{\If{$\exists$  decreasing interval-based operation}{
$op:=$ such an operation selected according to  {\bf Selection Rules}\;
}

}
\Else{
update clause weights according to the PAWS scheme\;
randomly sample $K$ interval-based operations into the set $Set_{op}$\;
$op:=$ the  best operation among $Set_{op}$ picked by {\bf Selection Rules}\; 
% Sample $K$ 
% \If{$\exists$ {\it cm} operation in $c$}{$op:=$ a {\it cm} operation in $c$  selected by {\bf Selection Rules}}
% \Else{$op:=$ assign a random variable in $c$ to a random value in its bound}
}
$\alpha:=\alpha$ with $op$ performed\; 
}
return ``UNKNOWN''\;

\end{algorithm}

\section{Experiments}
\label{experiments}
We conducted experiments to evaluate LocalSMT(RA) on 2 benchmark sets from SMT-LIB, and compare it with state-of-the-art SMT solvers and local search solvers.
Moreover, ablation experiments are conducted to analyze the effectiveness of the proposed strategies.

\subsection{Experiment Preliminaries}
{\bf Implementation: }
LocalSMT(RA) is implemented in C++ and compiled by g++ with '-O3' option.
There are 3 parameters in LS-LRA:
 $L$ for switching phases,  $K$ for the number of sampled operation and $sp$ (the smoothing probability) for the PAWS scheme.
The parameters are tuned according to suggestions from the literature~\cite{cai2022local,cai2013local} and our preliminary experiments on 20\% sampled instances.
Preliminary experiments show that LocalSMT(RA) is not sensitive to the parameter setting in a considerable range.
Parameters are set as follows:
$L=20$, $K=3$, $sp=0.0003$ for all benchmarks. 
In our implementation, to escape from local optimum, LocalSMT(RA) restarts every 500000 iterations.
We use a fixed value for the infinitesimal real number $\delta$, which is $min(\frac{1}{256}, \frac{1}{c_{max}})$, where $c_{max}$ denotes the maximum absolute value of coefficients in the formula.

Note that in contrast to previous local search for for SMT on Linear Integer Arithmetic~\cite{cai2022local}, our solver is able to handle arbitrary Boolean structure of formulas, including ``ite'' operator, by using the Tseitin encoding~\cite{prestwich2009cnf}.

Although any formula with high exponents can be rewritten as multi-linear formula by introducing fresh variables (for example, $x^2>4$ can be rewritten as $(x=y)\wedge(x*y>4)$), however, in that case, we cannot efficiently find the correct feasible solution of $x$ because of the relationship of both variables.
In our implementation, we simplify the formula by eliminating equations in form of $x = (a*y)$, where $x$ and $y$ are real-value variables and $a$ is coefficient.
Specifically, we will replace $x$ with $(a*y)$, and after simplifying, we only reserve those instances which are still multi-linear.

{\bf Competitors: }
In the context of SMT(LRA), we compare LocalSMT(RA) with the top 4 state-of-the-art SMT solvers according to  SMT-COMP 2022, namely
OpenSMT (version 2.4.2)
\footnote{https://github.com/usi-verification-and-security/opensmt},
Yices2 (version 2.6.2)
\footnote{https://yices.csl.sri.com},
cvc5 (version 1.0.0)
\footnote{https://cvc5.github.io/},
and Z3 (version 4.8.17)
\footnote{https://github.com/Z3Prover/z3/}.
While in the context of SMT(NRA), the top 4 competitors are as follows,
cvc5 (version 1.0.0),
Yices2 (version 2.6.2),
Z3 (version 4.8.17)
and SMT-RAT-MCSAT (version 22.06)
\footnote{https://github.com/ths-rwth/smtrat}.
The binaries of all competitors are downloaded from their websites.
Note that portfolio and derived solvers are excluded.

{\bf Benchmarks: } Experiments are carried out on 2 benchmark sets from SMT-LIB.
%LS-LIA is evaluated on 4 benchmarks in forms of SMT(LIA) and SMT(IDL).

\begin{itemize}
    \item SMTLIB-LRA: The benchmark set contains SMT(LRA) instances from SMT-LIB\footnote{https://clc-gitlab.cs.uiowa.edu:2443/SMT-LIB-benchmarks/QF\_LRA}.
    As LocalSMT(RA) is an incomplete solver, UNSAT instances are excluded, resulting in a benchmark with 1044  unknown and satisfiable instances. 
    \item SMTLIB-MRA: The benchmark set contains SMT(NRA) instances with multi-linear atomic constraints from  SMT-LIB\footnote{https://clc-gitlab.cs.uiowa.edu:2443/SMT-LIB-benchmarks/QF\_NRA}.
    UNSAT instances are also excluded, resulting in a benchmark with 979  unknown and satisfiable instances.
    
\end{itemize}

{\bf Experiment Setup: }
All experiments are carried out on a server with AMD EPYC 7763 64-Core  2.45GHz and 2048G RAM under the system Ubuntu 20.04.4.
Each solver is executed once with a cutoff time of 1200 seconds (as in the SMT-COMP) for each instance in both benchmark sets, as they contain sufficient instances (1044 for SMTLIB-LRA and 979 for SMTLIB-MRA).
We compare the number of instances where an algorithm finds a model (``\#solved''), as well as the run time.
Bold values in table emphasize the solver with greatest ``\#solved''.

% We  uploaded our solver in the anonymous Github repository\footnote{https://anonymous.4open.science/r/sls4lia/}.

\subsection{Results on SMTLIB-LRA Benchmark}

\subsubsection{Comparison with DPLL(T) solvers}
\begin{figure}
     \centering
     \begin{subfigure}[b]{0.45\textwidth}
         \centering
         \includegraphics[width=5.5cm]{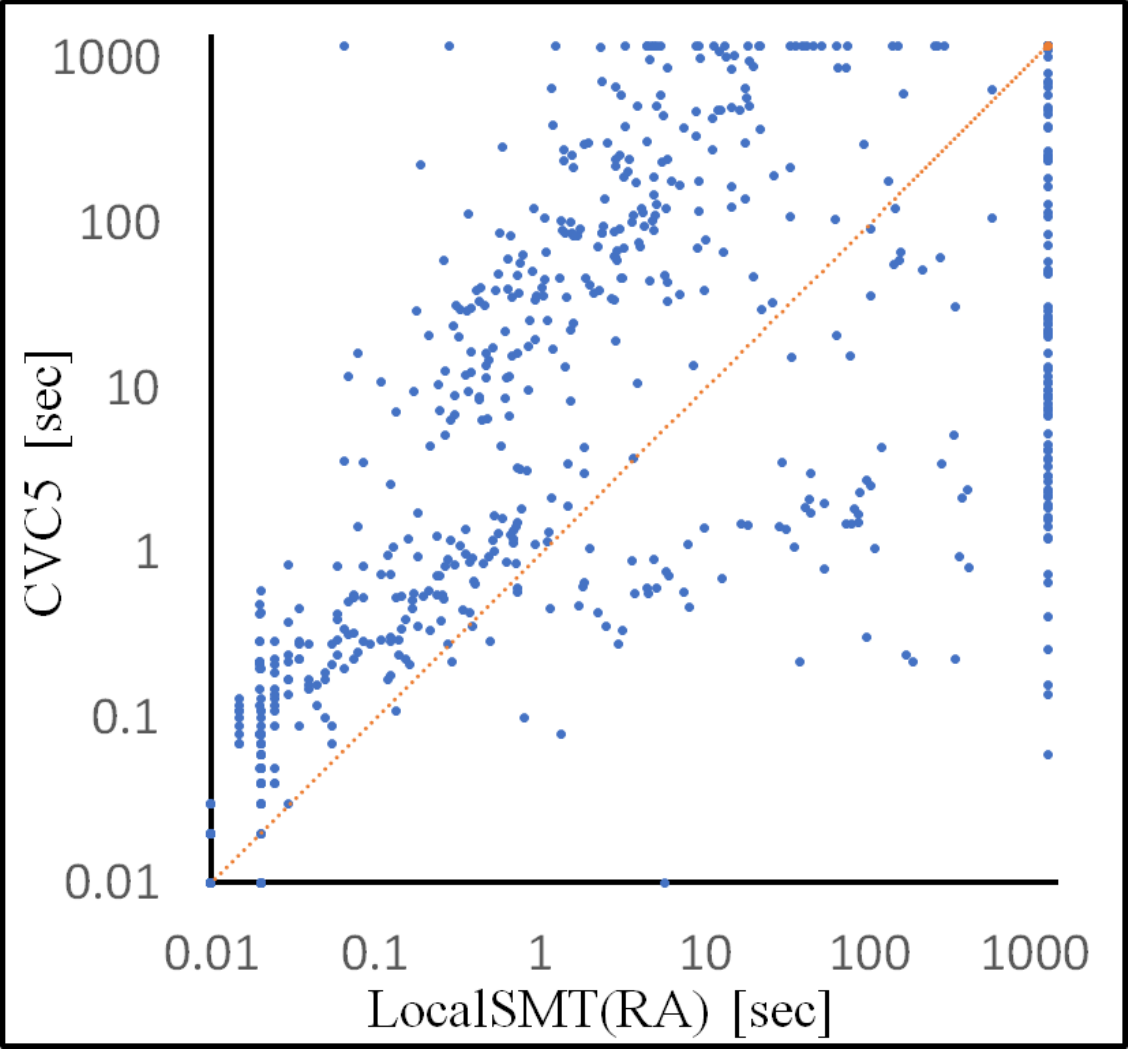}
         \caption{ Comparing with cvc5}
     \end{subfigure}
     \begin{subfigure}[b]{0.45\textwidth}
         \centering
         \includegraphics[width=5.5cm]{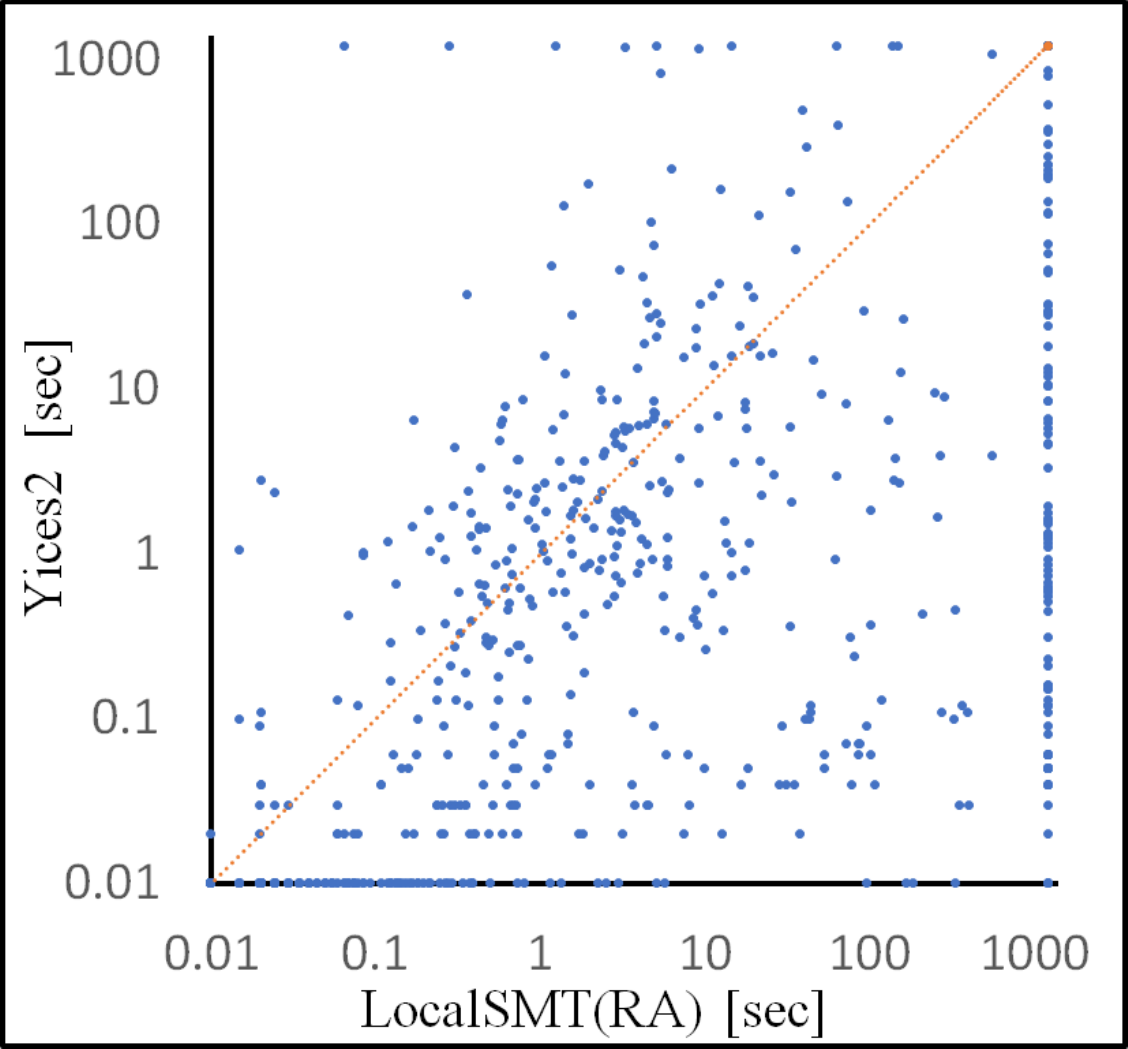}
         \caption{ Comparing with Yices2}
     \end{subfigure}

     \begin{subfigure}[b]{0.45\textwidth}
         \centering
         \includegraphics[width=5.5cm]{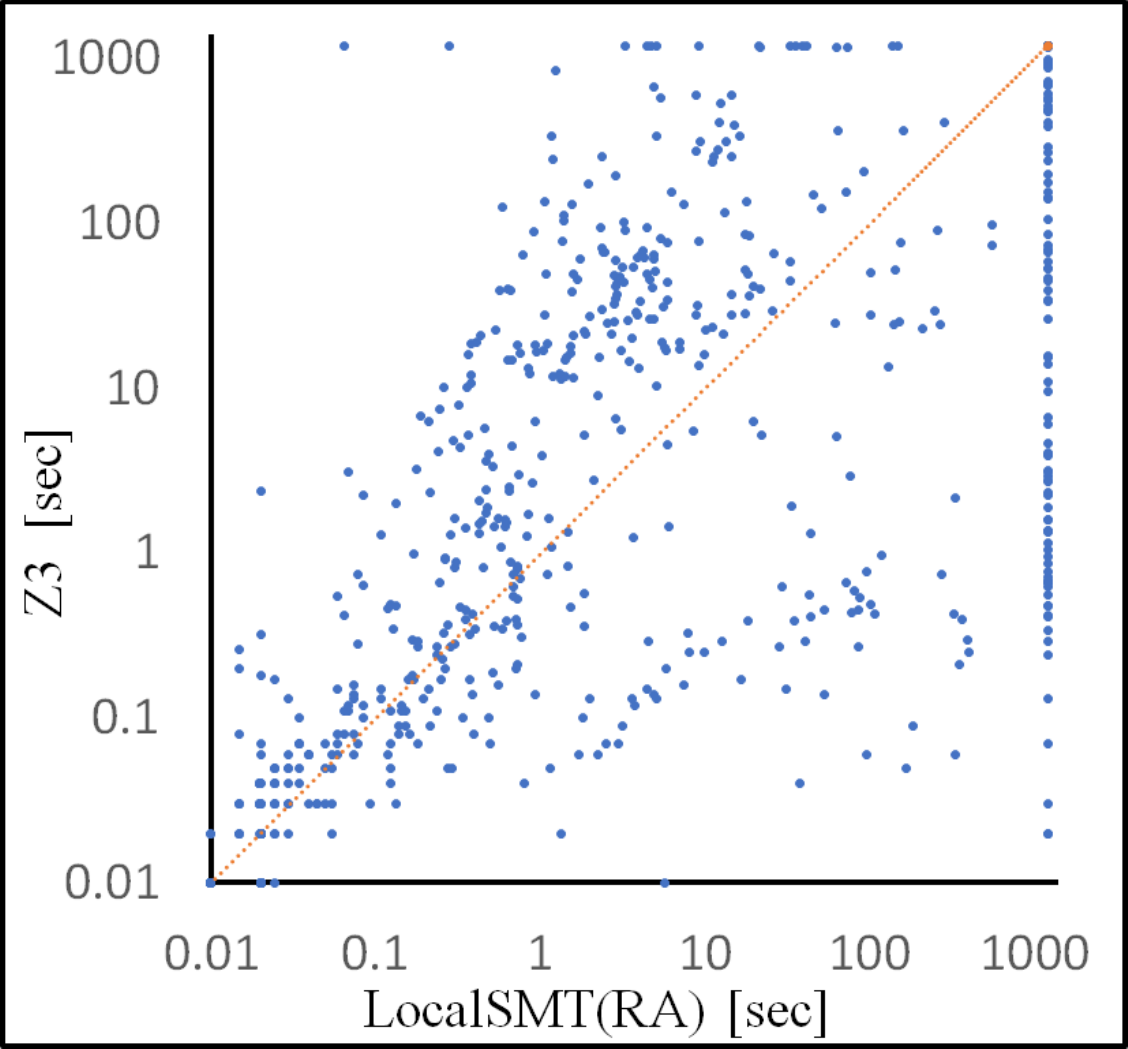}
         \caption{ Comparing with Z3}
     \end{subfigure}
     \begin{subfigure}[b]{0.45\textwidth}
         \centering
         \includegraphics[width=5.5cm]{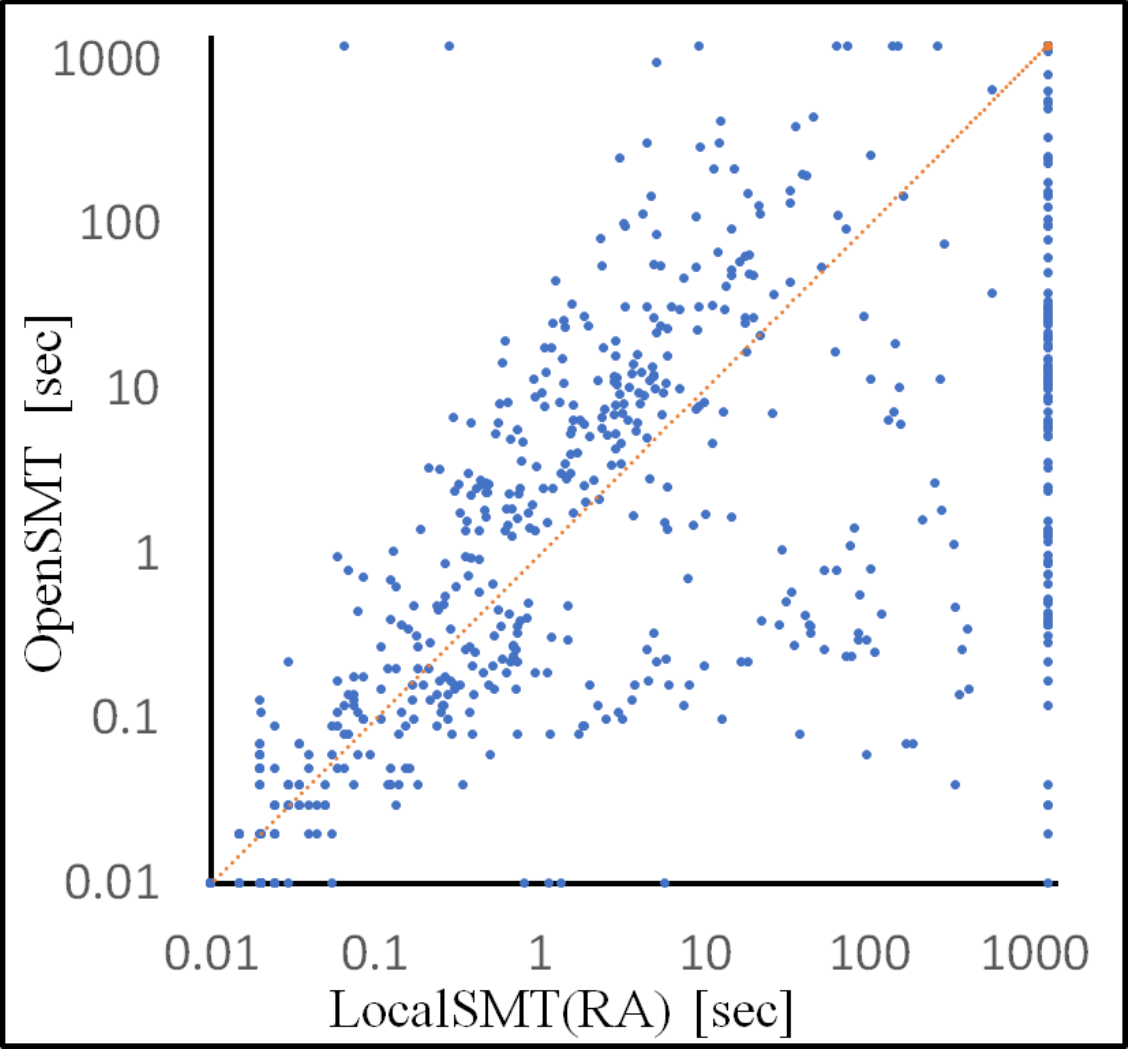}
         \caption{ Comparing with OpenSMT}
     \end{subfigure}
        \caption{Run time comparison on instances from SMTLIB-LRA}
        \label{fig:lra}
\end{figure}

As shown in Table \ref{table_lra}, LocalSMT(RA) can solve 900 out of 1044 instances, which is competitive but still cannot rival its competitors.
We also perform a runtime comparison between LocalSMT(RA) and each competitor on instances from SMTLIB-LRA in Fig \ref{fig:lra}, which shows that LocalSMT(RA) is complementary to the competitors.

One explanation for the fact that LocalSMT(RA) cannot rival its DPLL(T) competitors is that 
54.5\% of the instances in SMTLIB-LRA contain Boolean variables, while the Boolean mode of LocalSMT(RA) is not good at exploiting the relations among Boolean variables, similar to previous local search for SMT on Linear Integer Arithmetic~\cite{cai2022local}.
Another possible reason accounting for the poor performance of LocalSMT(RA) on this benchmark set is that our algorithm is not complete in the sense of probabilistically approximately complete (PAC), because the candidate value selection of {\it interval-based operation} can miss the possible satisfying solution where the values of variables are not considered.

\begin{table}[]
\centering
\caption{Results on instances from SMTLIB-LRA}
\setlength{\tabcolsep}{2.1mm}\scalebox{1.0}{
\begin{tabular}{@{}lllllll@{}}
\toprule
                                         & \#inst & cvc5        & Yices        & Z3          & OpenSMT     & LocalSMT(RA)     \\ \midrule
2017-Heizmann & 8      & 4           & 3            & 4           & 4           & \textbf{7} \\
2019-ezsmt                               & 84     & 61          & 61           & 53          & \textbf{62} & 35         \\
check                                    & 1      & 1           & 1            & 1           & 1           & 1          \\
DTP-Scheduling                           & 91     & 91          & 91           & 91          & 91          & 91         \\
LassoRanker                              & 271    & 232         & \textbf{265} & 256         & 262         & 240        \\
latendresse                              & 16     & 9           & \textbf{12}  & 1           & 10          & 0          \\
meti-tarski                              & 338    & 338         & 338          & 338         & 338         & 338        \\
miplib                                   & 22     & 14          & \textbf{15}  & \textbf{15} & \textbf{15} & 4          \\
sal                                      & 11     & 11          & 11           & 11          & 11          & 11         \\
sc                                       & 108    & 108         & 108          & 108         & 108         & 108        \\
TM                                       & 24     & \textbf{24} & \textbf{24}  & \textbf{24} & \textbf{24} & 11         \\
tropical-matrix                          & 10     & 1           & \textbf{6}   & 4           & \textbf{6}  & 0          \\
tta                             & 24     & 24          & 24           & 24          & 24          & 24         \\
uart                                     & 36     & \textbf{36} & \textbf{36}  & \textbf{36} & \textbf{36} & 30         \\
                                         &        &             &              &             &             &            \\
Total                                    & 1044   & 954         & \textbf{995} & 966         & 992         & 900        \\ \bottomrule
\end{tabular}
}
\label{table_lra}
\end{table}

\subsubsection{Comparison with Local Search Solvers}
We compare LocalSMT(RA) with a previous local search solver dedicated for SMT(LRA)~\cite{griggio2011stochastic}, called WalkSMT.
The best two configurations of WalkSMT, namely ``WalkSMT UBCSAT'' and ``WalkSMT UBCSAT++'', are adopted for comparison.
Since WalkSMT only supports the earlier version of SMT-LIB standard, which has been deprecated, we perform experiments using the same experiment setup in their paper~\cite{griggio2011stochastic}, where the cutoff is set as 600s, and we compare with the result presented in the original paper.
The results are shown in Table \ref{tbl:sls}, indicating that LocalSMT(RA) can significantly outperform both configurations of WalkSMT, especially on the ``sc'' type.

\begin{table}[]
\caption{Comparison with WalkSMT on instances from SMTLIB-LRA}
\begin{tabular}{@{}llllllll@{}}
\toprule
                 & sc           & uart        & sal         & TM          & tta         & miplib     & Total        \\ \midrule
\#inst           & 108          & 36          & 11          & 24          & 24          & 22         & 225          \\
WalkSMT UBCSAT   & 78           & \textbf{35} & \textbf{10} & 14          & 9           & \textbf{3} & 149          \\
WalkSMT UBCSAT++ & 63           & 14          & 8           & \textbf{19} & 10          & 2          & 116          \\
LocalSMT(RA)            & \textbf{106} & 29          & 8           & 9           & \textbf{24} & \textbf{3} & \textbf{179} \\ \bottomrule
\end{tabular}
\label{tbl:sls}
\end{table}

% \subsection{Results on multi-linear instances from SMTLIB-NRA Benchmark}
\subsection{Results on SMTLIB-MRA Benchmark}
\begin{figure}
     \centering
     \begin{subfigure}[b]{0.45\textwidth}
         \centering
         \includegraphics[width=5.5cm]{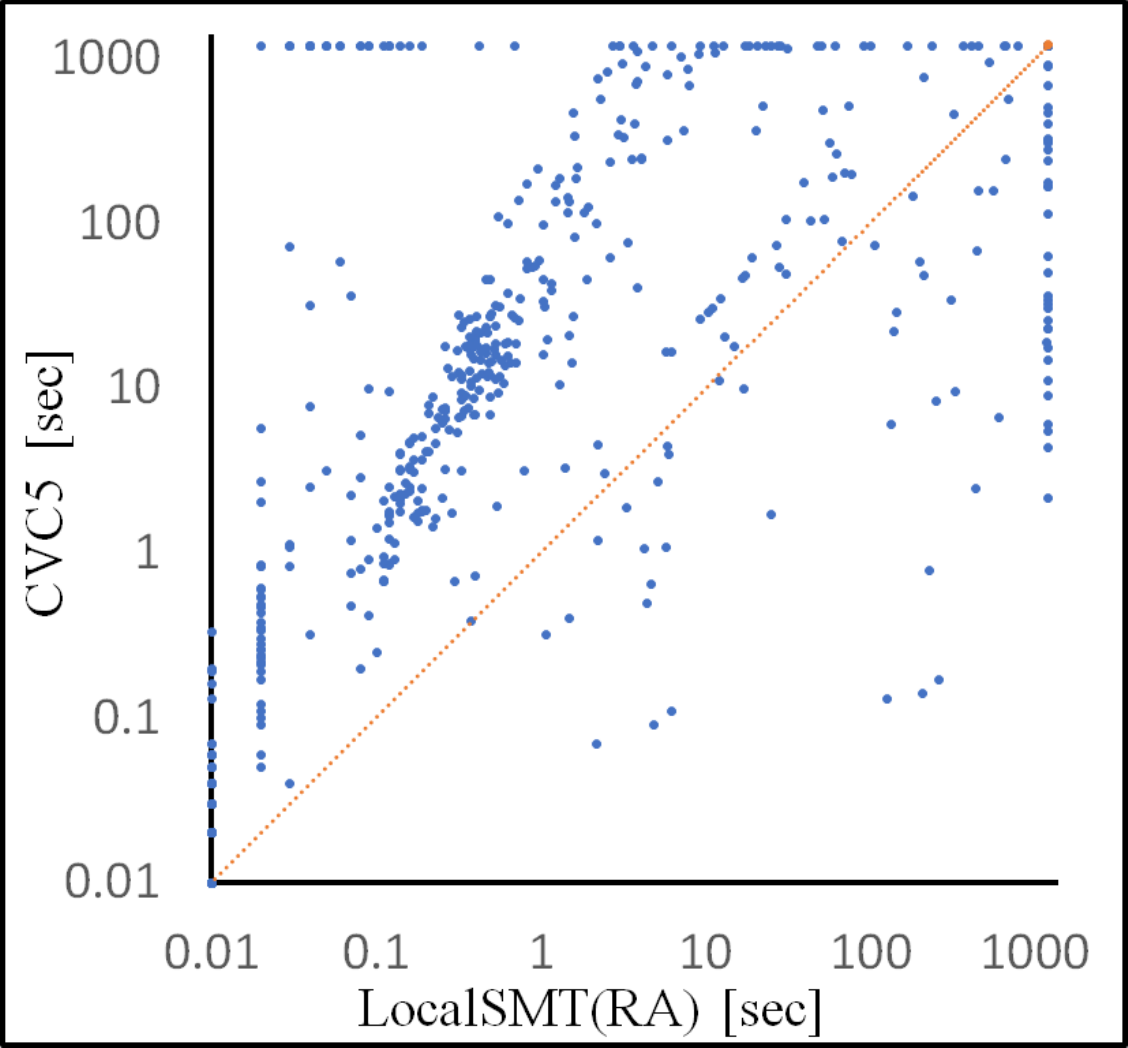}
         \caption{ Comparing with cvc5}
     \end{subfigure}
     \begin{subfigure}[b]{0.45\textwidth}
         \centering
         \includegraphics[width=5.5cm]{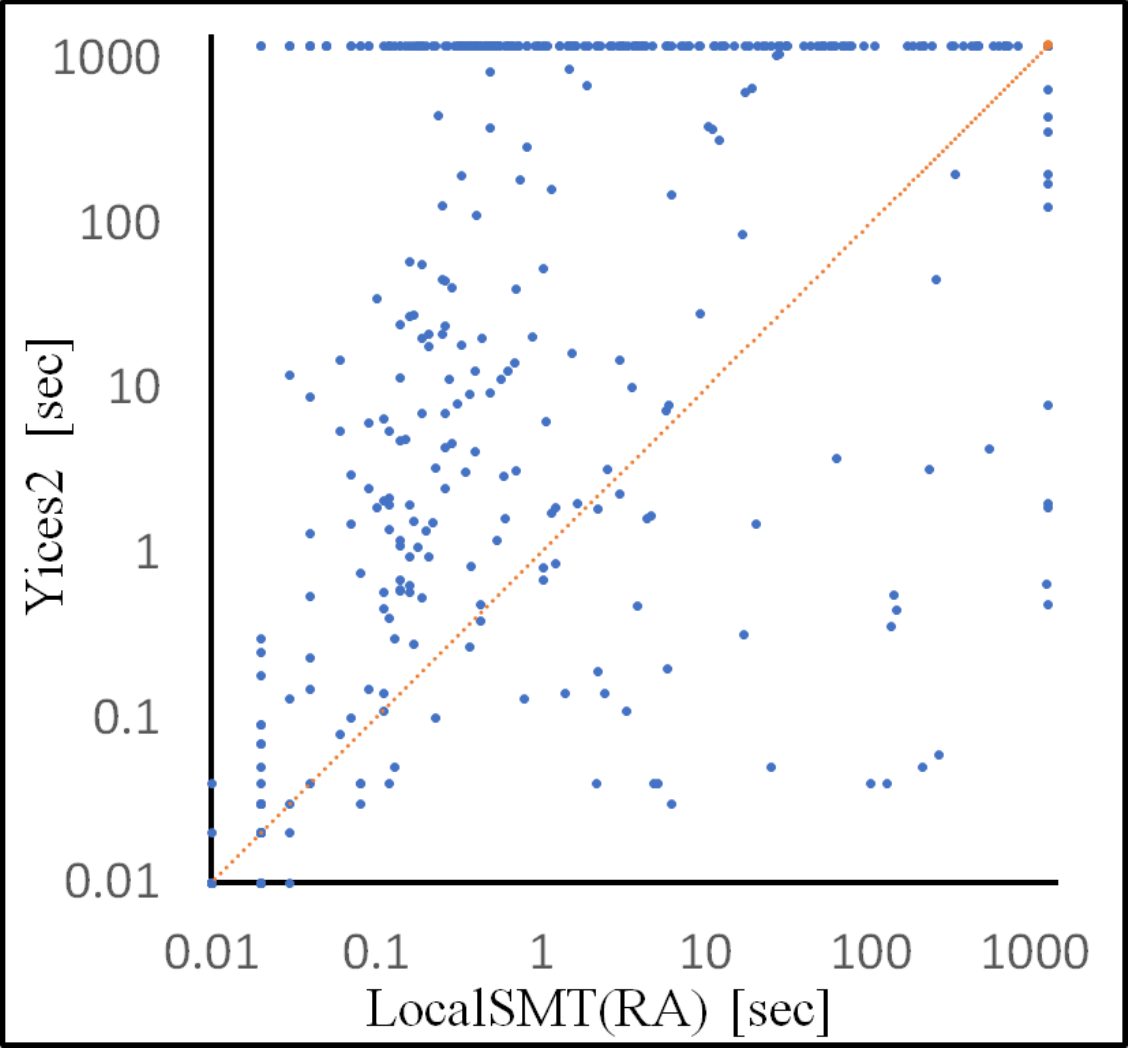}
         \caption{ Comparing with Yices2}
     \end{subfigure}

     \begin{subfigure}[b]{0.45\textwidth}
         \centering
         \includegraphics[width=5.5cm]{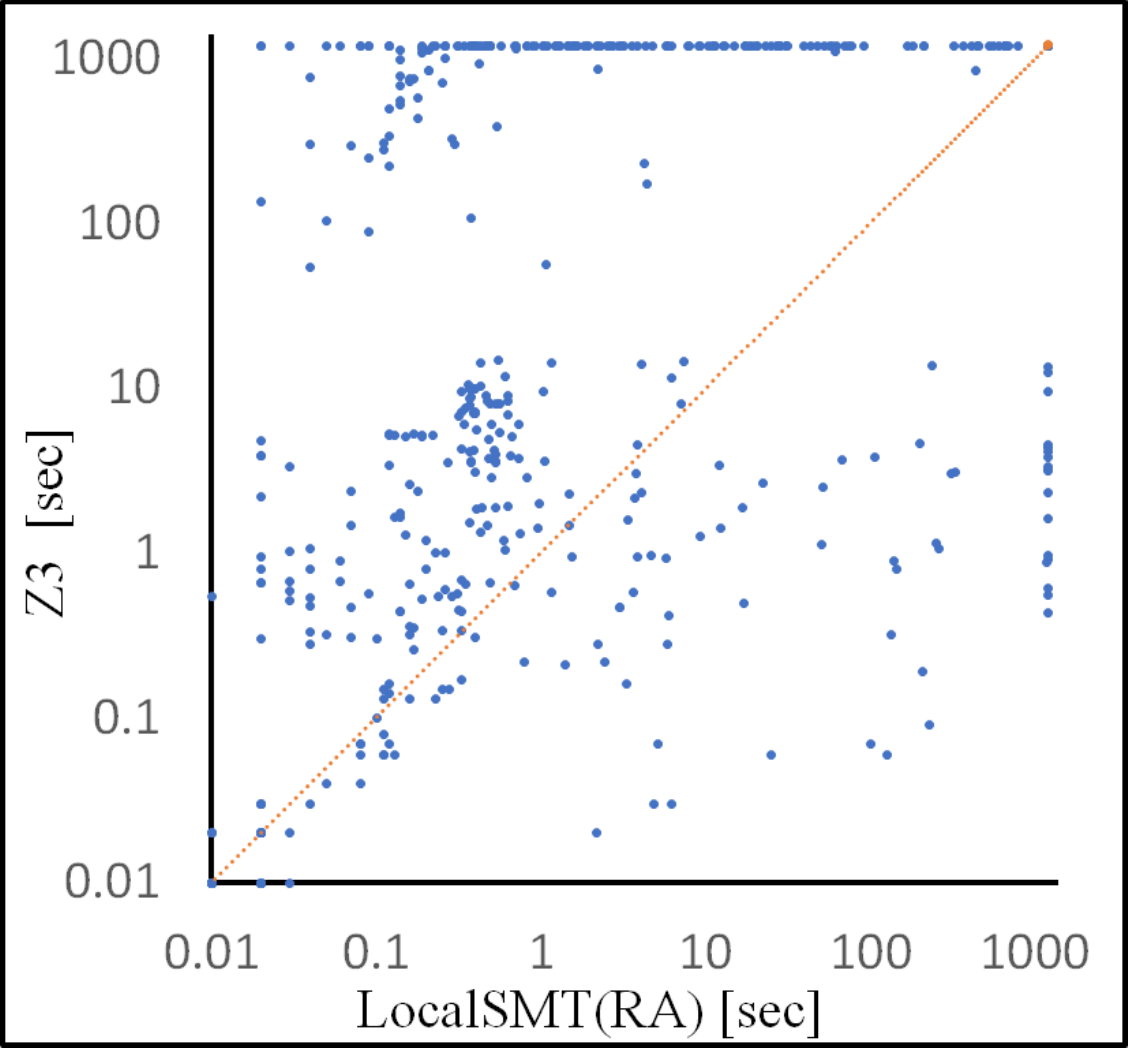}
         \caption{ Comparing with Z3}
     \end{subfigure}
     \begin{subfigure}[b]{0.45\textwidth}
        \centering
         \includegraphics[width=5.5cm]{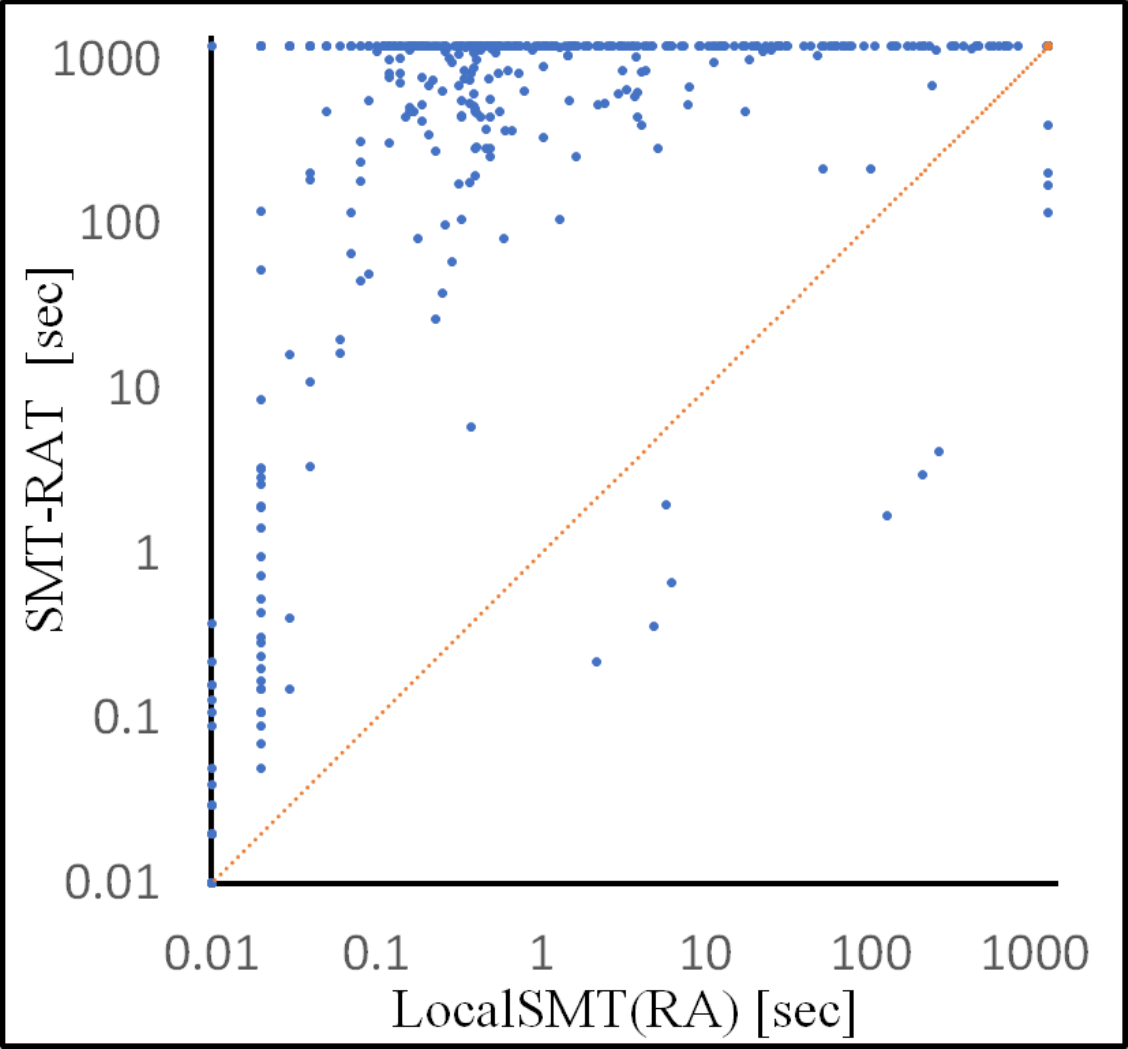}
         \caption{ Comparing with SMT-RAT}
     \end{subfigure}
        \caption{Run time comparison on instances from SMTLIB-MRA}
        \label{fig:nra}
\end{figure}

LocalSMT(RA) can solve more multi-linear instances than all competitors on this benchmark set (solving 891 out of 979 instances), which is shown in Table \ref{table_nra}. 
Moreover, LocalSMT(RA) can uniquely solve 28 instances in this benchmark set.
The time comparison between LocalSMT(RA) and its competitors is shown in Fig \ref{fig:nra}, confirming that LocalSMT(RA) is more efficient than all competitors in SMTLIB-MRA.
As shown in Table \ref{small time}, LocalSMT(RA) shows better performance with smaller cutoff.
Specifically, LocalSMT(RA) can solve 138, 122 and 99 more instances than the best of competitors respectively with the cutoff of 1s, 5s, and 10s.

LocalSMT(RA) works particularly well on instances from ``zankl'' and ``UltimateAutomizer'' type, which are industrial instances generated in the context of software verification.
On these types, LocalSMT(RA) can solve all instances, outperforming all the competitors.
Moreover, LocalSMT(RA) can exclusively solve 13 instances from ``20170501-Heizmann'' type, which implements a constraint-based synthesis of invariants\cite{colon2003linear}.

The explanation for the superiority of LocalSMT(RA) on SMTLIB-MRA are as follows.
In contrast to LRA, the theory solver for NRA constraints requires complex calculation, which reduces the performance of these competitors, while LocalSMT(RA) can trivially determine the operations in multi-linear constraints, and thus LocalSMT(RA) can efficiently explore the search space.
Moreover, In SMTLIB-MRA benchmarks, 3.4\% instances have Boolean variables. In contrast to SMTLIB-LRA, whose counterpart is 54.5\%, SMTLIB-MRA has simpler Boolean structures.

\begin{table}[]
\centering
\caption{Results on instances from SMTLIB-MRA}
\setlength{\tabcolsep}{1.9mm}\scalebox{1.0}{
\begin{tabular}{@{}lllllll@{}}
\toprule
                                             & \#inst & cvc5         & Yices       & Z3         & SMT-RAT          & LocalSMT(RA)       \\ \midrule
20170501-Heizmann & 51     & 1            & 0           & 4          & 0           & \textbf{17}  \\
20180501-Economics                  & 28     & 28           & 28          & 28         & 28          & 28           \\
2019-ezsmt                                   & 32     & 31           & \textbf{32} & \textbf{32}& 21 & 28           \\
20220314-Uncu                                & 12     & 12           & 12          & 12         & 12          & 12           \\
LassoRanker                                  & 347    & \textbf{312} & 124         & 199        & 0         & 297          \\
meti-tarski                                  & 423    & 423          & 423         & 423        & 423         & 423          \\
UltimateAutomizer                            & 48     & 34           & 39          & 46         & 18          & \textbf{48}  \\
zankl                                        & 38     & 24           & 25          & 28         & 30          & \textbf{38}  \\ 
                                             &        &              &             &            &             &            \\
Total                                        & 979    & 865          & 683         & 772        & 532         & \textbf{891} \\ \bottomrule
\end{tabular}
}
\label{table_nra}
\end{table}

\begin{table}[]
\centering
\caption{Results on instances from SMTLIB-MRA with different cutoff}
\label{small time}
\begin{tabular}{@{}ccccccc@{}}
\toprule
Cutoff &    & cvc5 & Yices & Z3  & SMT-RAT & LocalSMT(RA) \\ \midrule
1s  & & 535  & 567   & 595 & 495     & \bf{733}   \\
5s  & & 607  & 609   & 674 & 507     & \bf{796}   \\
10s & & 649  & 623   & 714 & 508     & \bf{813}   \\ \bottomrule
\end{tabular}
\end{table}

\subsection{Effectiveness of Proposed strategies}
To analyze the effectiveness of the strategies in LocalSMT(RA), we modify  LocalSMT(RA) to obtain 3 alternative versions as follows.
\begin{itemize}
    \item To analyze the effectiveness of {\it interval-based} operator, LocalSMT(RA) is modified by replacing the operator with traditional $cm$ operator, leading to the version $v\_{cm}$.
    \item To analyze the effectiveness of the {\it tie-breaking mechanism}, we modify LocalSMT(RA) by evaluating operation based on $score$ without considering the {\bf Selection Rules}, that is to randomly pick one operation with greatest $score$, leading to the version $v\_{score}$.
    \item We also implement a plain version which adopts neither {\it interval-based} operator nor {\it tie-breaking mechanism}, denoted as $v\_{plain}$.
\end{itemize}

\begin{figure}[t]
     \centering
     \begin{subfigure}[b]{0.49\textwidth}
         \centering
         \includegraphics[width=8cm]{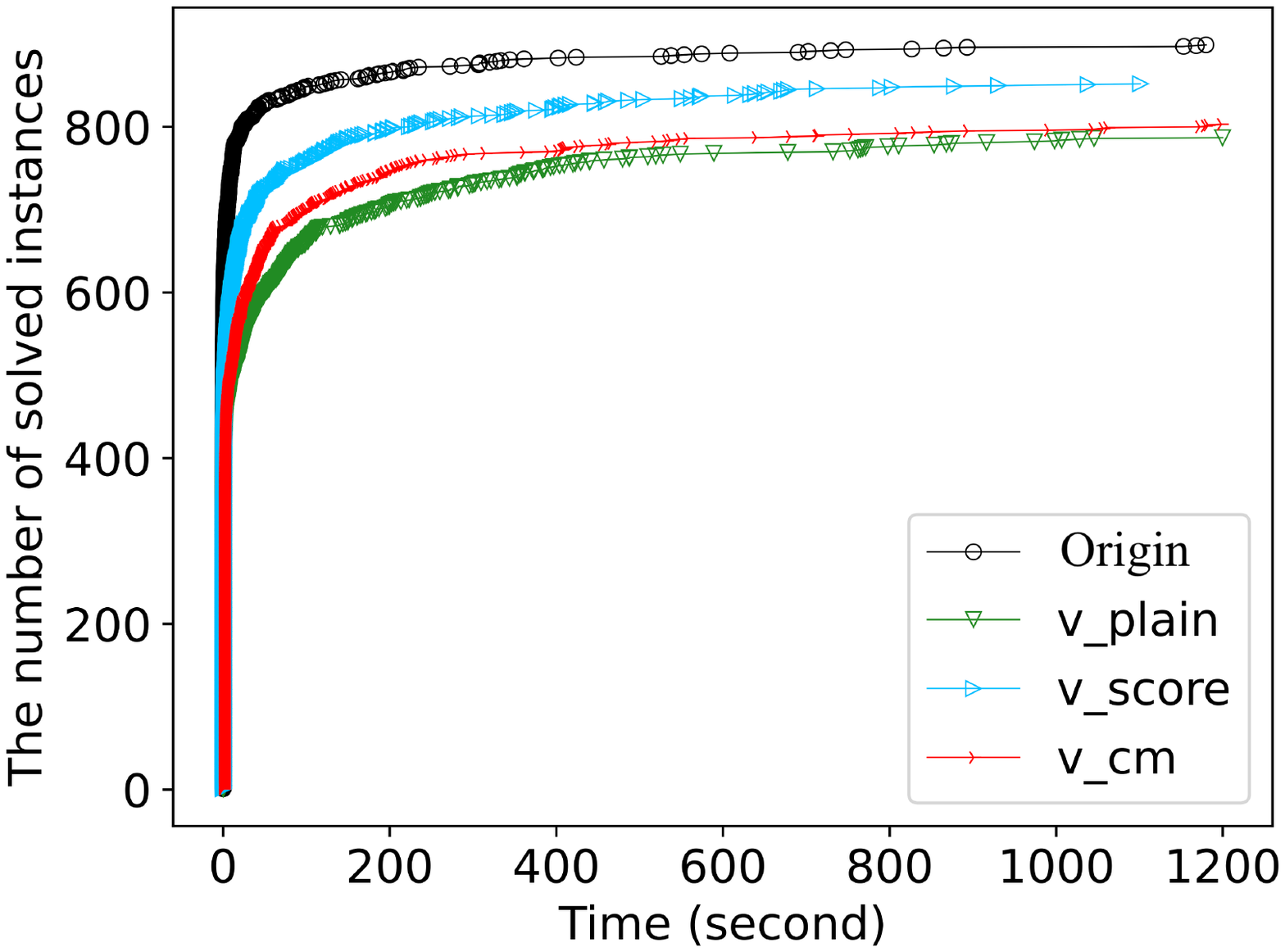}
         \caption{Comparison on SMTLIB-LRA}
     \end{subfigure}
     \begin{subfigure}[b]{0.49\textwidth}
         \centering
         \includegraphics[width=8cm]{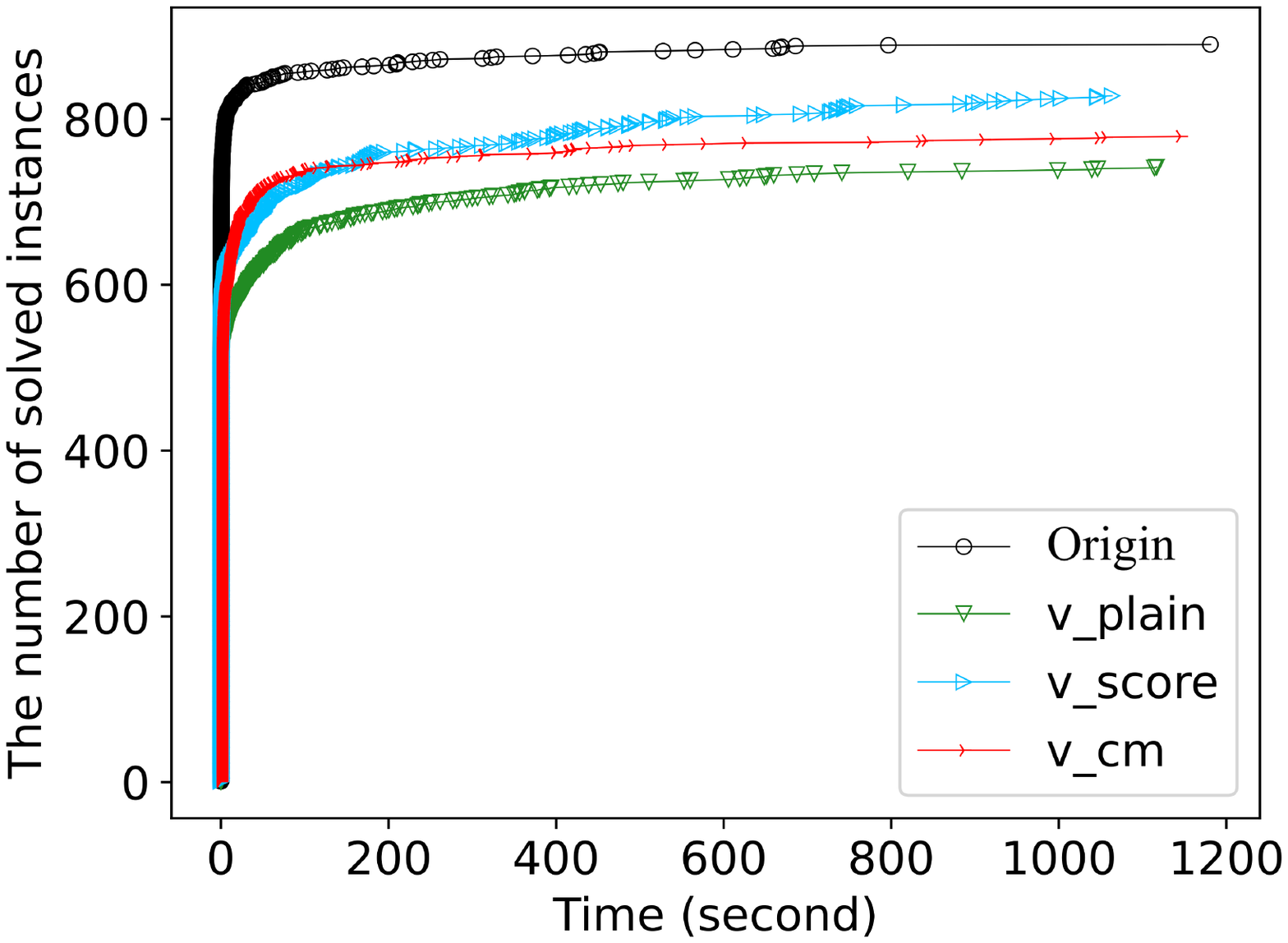}
         \caption{Comparison on SMTLIB-MRA}
     \end{subfigure}
        \caption{Run time distribution comparison}
        \label{distribution}
\end{figure}

The original version of LocalSMT(RA), denoted as $Origin$, is compared with these modified versions on both benchmarks.
The runtime distribution of LocalSMT(RA) and its modified versions is presented in Fig. \ref{distribution}, which confirms the effectiveness of our proposed strategies.

\section{Conclusion and Future Work}
\label{conclusion}
In this paper, we propose the first local search algorithm for SMT on Real Arithmetic based on the following novel ideas.
First, the {\it interval-based} operation is proposed to enrich the traditional critical move operator, by considering the interval information.
Moreover, a {\it tie-breaking} mechanism is proposed to distinguish  operations with same best $score$. 
Experiments on SMT-LIB show that our solver is competitive and complementary with state-of-the-art SMT solvers, especially for those multi-linear instances.

The future direction is to extend LocalSMT(RA) to support all SMT(NRA) instances and deeply combine our local search algorithm with the  DPLL(T) SMT solver, resulting in a hybrid solver that can make the most of respective advantages.
Moreover, we will enrich the sampled candidate values of {\it interval-based} operation by considering more random values with small denominator.

\section{Acknowledgements}{This work is supported by NSFC Grant 62122078.}



\bibliographystyle{IEEEtran}
\bibliography{a}


% \end{thebibliography}

\end{document}